# A new method for analyzing the collimation angle of a neutron Soller collimator[1]

GAO Jian-Bo (高建波)[1)]  LIU Yun-Tao（刘蕴韬）[2)*]  CHEN Dong-Feng（陈东风）[3)*]

China Institute of Atomic Energy, Beijing, China, 102413

**Abstract.** A new method for analyzing the collimation angle of a neutron Soller collimator is described. A Gaussian distribution formula is used to define the angular distribution function of the neutron source and the neutron transmission function of the Soller collimator. A relationship between the FWHM of the collimator rocking curve and the collimation angle is derived. Using this method, some rocking curve experiment results are analyzed. The results show that the new function can be a good theoretical model for fitting the experimental data, especially for the data of two collimators with different collimation angles.

**PACS**：29.27.Eg

## 1. Introduction

In neutron and X-ray scattering experiments, Soller slit collimators are extensively used to obtain radiation beams with a given collimation angular and hence to optimize the instrumental resolution [1-4]. A conventional neutron Soller collimator usually consists of many thin, parallel absorbing plates separated by spacers in a supporting frame. The absorbers are assumed to be of thickness $t$, with separation of the absorbing plates (center to center) $S$ and length $L$. Then the collimation angle $\Gamma$ is defined by $tan^{-1}[(S-t)/L]$. The ideal theoretical neutron beam transmission expression for a Soller collimator has a triangular form as a function of angular divergence. The full-width at half-maximum (FWHM) of the triangular profile is just the same value of collimation angle $\Gamma$ while the full-width at the base of the triangle is twice this value.

Meister and Weckermann gave a theoretical five-order polynomial expression to describe the rocking curve function for two identical Soller collimators [5]. In this article a new theoretical Gaussian expression to describe the rocking curve function is introduced.

## 2. Theory

The setup for the collimation angle measuring process is shown in Figure 1. A neutron beam coming from a neutron source passes through measuring collimator 1 and reference collimator 2 and finally hits the detector. While rocking collimator 1, the detector records the neutron intensity as a function of rocking angle. When the experiment is finished, a theoretical expression is used to fit the rocking curve data. The collimation angle is then obtained by analyzing the fitted parameters.

Received 19 March 2015
1 Supported by National Natural Science Foundation of China (11205248)

1)email: jianbo@ciae.ac.cn

2)*Corresponding author: dongfeng@ciae.ac.cn

3)*Corresponding author: ytliu@ciae.ac.cn

Meister and Weckermann introduced a method for analyzing the collimation angle of a neutron Soller collimator [5], and until now many experiment results have been analyzed using this method [6-8]. A mathematical model for the neutron source and neutron transmission of a collimator was given according to the experiment process. A parabolic expression was used to describe the angular distribution function of the neutron source, and a triangular expression was used to describe the neutron transmission function. After performing a convolution integral for the neutron source and collimator, the theoretical rocking curve function was derived to be a five-order polynomial expression. By fitting the experiment data, the collimation angle was then obtained.

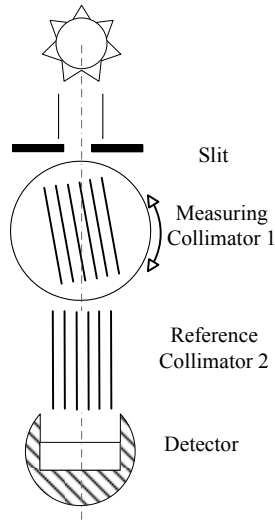

Fig.1. Arrangement of collimator measurement

In this article a new theoretical expression for describing the rocking curve function is introduced: a Gaussian distribution formula is used to describe the angular distribution function of the neutron source. The ideal theoretical function for the collimator transmission is the triangle function. In practical application the absorbing plates of the collimator might not absorb all the neutrons perfectly and so some neutrons could pass through, while other neutrons might be scattered in any direction after hitting the collimator. Taking these reasons into account, the Gaussian function could be a good approximation for neutron transmission. After a convolution integral for Gaussian distribution formulas, the theoretical rocking curve function is derived to still be a Gaussian distribution expression. A relationship between the FWHM of the rocking curve and the collimation angle is given by a relatively simple expression. The details of the derivation are as follows.

The angular distribution function of the neutron source is defined by a Gaussian distribution formula

$$I_s(x) = I_0 \exp(-\frac{x^2}{2\sigma_s^2}), \qquad (1)$$

where x is the angle between the neutron incidence and the beam axis, and can be an arbitrary real number. The angular distribution function of neutron transmission for a Soller collimator is defined by the Gaussian distribution formula

$$T(x) = T \exp(-\frac{x^2}{2\sigma^2}). \qquad (2)$$

The rocking curve function can be convolution integrated by $I_S(x) \cdot T_2(x)$ and $T_1(x)$,

$$I_{s+1+2}(x)=\int I_s(\gamma)T_1(x-\gamma)T_2(\gamma)d\gamma=A\exp(-\frac{x^2}{2\sigma_{s+1+2}^2}), \tag{3}$$

where

$$\sigma_{s+1+2}^2 = 1/(1/\sigma_s^2 +1/\sigma_2^2)+\sigma_1^2. \tag{4}$$

For the Gaussian distribution function, we know that

$$\Gamma=FWHM=2.355\sigma, \tag{5}$$

so we can get

$$\Gamma_{s+1+2}^2=1/(1/\Gamma_s^2+1/\Gamma_2^2)+\Gamma_1^2. \tag{6}$$

## 3. Discussion

Table 1. Comparison of theoretical models presented in this article and reference [5].

| | This article | Reference [5] |
|---|---|---|
| Neutron source | $I_s(x)=I_0\exp(-\frac{x^2}{2\sigma_s^2})$ | $I_s(x)=I_0(1-cx^2)$, $\|x\|\leq 1/\sqrt{c}$ |
| Neutron transmission | $T(x)=T\exp(-\frac{x^2}{2\sigma^2})$ | $T(x)=\begin{cases} Tx/\Gamma+T & (-\Gamma\leq x\leq 0) \\ -Tx/\Gamma+T & (0\leq x\leq\Gamma) \\ 0 & (x\leq-\Gamma, x\geq\Gamma) \end{cases}$ |
| Neutron rocking curve | $I_{s+1+2}(x)=A\exp(-\frac{x^2}{2\sigma_{s+1+2}^2})$ | $[I_{s+1+2}(x)]_{-2\Gamma}^{-\Gamma}$, $-2\Gamma\leq x\leq-\Gamma$ |
| | | $[I_{s+1+2}(x)]_{-\Gamma}^{0}$, $-\Gamma\leq x\leq 0$ |

$I_{s+1+2}(x)$ from reference [5] is symmetrical to x=0 and is zero for $2\Gamma \leq |x|$.

$$[I_{s+1+2}(x)]_{-2\Gamma}^{-\Gamma} = I_0(T/\Gamma)^2[4\Gamma^3/3 - 4c\Gamma^5/15$$
$$+x(2\Gamma^2-2c\Gamma^4/3)+x^2(\Gamma-c\Gamma^3)$$
$$+x^3(1/6-5c\Gamma^2/6)-x^4c\Gamma/3-x^5c/20] \tag{7}$$

for the angular range $-2\Gamma \leq x \leq -\Gamma$, and

$$[I_{s+1+2}(x)]_{-\Gamma}^{0} = I_0(T/\Gamma)^2[2\Gamma^3/3 - c\Gamma^5/15 - x^2\Gamma$$
$$-x^3(1/2-c\Gamma^2/6)+x^4c\Gamma/3+x^5 3c/20] \tag{8}$$

for the angular range $-\Gamma \leq x \leq 0$.

Table 1 shows a comparison of the theoretical models presented in this article and in Reference [5]. Meister and Weckermann used a parabolic expression to describe the angular distribution function of the neutron source. Because the neutron intensity cannot be negative, the angular dispersion x should be set within a reasonable range. Hence, the FWHM of the neutron source is limited by FWHMs ≥1.41 • Γ. This means that the FWHM of the neutron source should be a bit larger than the collimation angle of the testing collimator when this model is applied. This is an implied precondition for Meister and Weckermann's model. Another precondition one should keep in mind for the reference model is that the collimation angle of the reference collimator should be exactly the same as that of the testing one. For the new model in this article, however, there are no such limitations. The FWHM of the neutron source can be smaller and the collimation angle of the reference collimator can be different using the new model, without causing difficulties for analysis.

Another advantage of the Gaussian model is that the FWHM of the rocking curve has a relatively simple relationship with collimation angles from the neutron source, testing collimator and reference collimator. By using Eq. (6), one can easily estimate the collimation angle with the corresponding error. Generally the FWHM of the neutron source is larger than that of the collimator. Considering the collimation angle of the testing collimator is the same as that of the reference collimator, if the FWHM of the neutron source is three times larger than that of the reference collimator, the FWHM of the rocking curve can be reduced to

$$\Gamma_{1+2}^2 = \Gamma_2^2 + \Gamma_1^2. \tag{9}$$

The relative deviation is 2.60% compared with the unreduced result. If the FWHM of the neutron source is more than five times larger than the reference collimator, the relative deviation between the two results is less than 1%. Some other results are shown in Table 2.

Table 2. FWHM of the rocking curve with different collimation angles of neutron source

| $\Gamma_s$ (′) | $\Gamma_1$ (′) | $\Gamma_2$ (′) | $\Gamma_s/\Gamma_1$ | $\Gamma_{s+1+2}$ (′) | $\Gamma_{1+2}$ (′) | Relative Deviation |
|---|---|---|---|---|---|---|
| 20 | 10 | 10 | 2 | 13.42 | 14.14 | 5.41% |
| 30 | 10 | 10 | 3 | 13.78 | 14.14 | 2.60% |
| 40 | 10 | 10 | 4 | 13.93 | 14.14 | 1.50% |
| 50 | 10 | 10 | 5 | 14.01 | 14.14 | 0.98% |
| 60 | 10 | 10 | 6 | 14.05 | 14.14 | 0.68% |

## 4. Error analysis

If the FWHM of the neutron source is much more than that of the collimators, the rocking curve of the two collimators is the convolution of two triangular profile functions, according to the model from Meister and Weckermann. The expression is,

$$I(x) = I_0 T^2 (4\Gamma/3 + 2x + x^2/\Gamma + x^3/6\Gamma^2) \qquad -2\Gamma \le x \le -\Gamma. \tag{10}$$

$$I(x) = I_0 T^2 (2\Gamma/3 - x^2/\Gamma - x^3/2\Gamma^2) \qquad -\Gamma \le x \le 0. \tag{11}$$

The FWHM from the numerical calculation is $1.44\Gamma$. By using the new model in this article, the rocking curve of two Gaussian functions is still a Gaussian function, and the FWHM is $\sqrt{2}\Gamma$. The relative deviation between the two results is 1.81%.

Using the experiment parameters $\Gamma=2.94\times10^{-3}$ and $c=1.8\times10^4$ from Meister and Weckermann's article, the FWHM of the rocking curve is fitted to be 14.43′. By using the new method, a Gaussian model with the same FWHM (36.24′) as the parabolic model is used to describe the neutron source. The new FWHM of the rocking curve is calculated to be 14.03′. The difference between the two FWHMs is 2.77%.

Another article introduced a FWHM result fitted by using Meister and Weckermann's method[6]. The parameter $\Gamma$ was 10.72′, and c was $1.34\times10^4$. The FWHM of the neutron source was 42.00′. By using Meister and Weckermann's method, the FWHM of the rocking curve is fitted to be 15.35′, while by using the new method the FWHM is calculated to be 14.93′. The relative deviation between the two results is 2.73%.

Table 3. Calculated FWHM of the rocking curve by new method ($\Gamma_{s+1+2\text{-new}}$) compared to method from Ref. [5] ($\Gamma_{s+1+2\text{-ref}}$)

| $\Gamma_s$ (′) | $\Gamma_1$ (′) | $\Gamma_2$ (′) | $\Gamma_{s+1+2\text{-new}}$ (′) | $\Gamma_{s+1+2\text{-ref}}$ (′) | Relative Deviation |
|---|---|---|---|---|---|
| 50 | 10 | 10 | 14.01 | 14.36 | 2.44% |
| 60 | 10 | 10 | 14.05 | 14.39 | 2.36% |
| 70 | 20 | 20 | 27.75 | 28.55 | 2.80% |
| 80 | 20 | 20 | 27.87 | 28.63 | 2.65% |
| 90 | 10 | 10 | 14.10 | 14.42 | 2.22% |
| 90 | 20 | 20 | 27.95 | 28.69 | 2.58% |
| 90 | 30 | 30 | 41.35 | 42.63 | 3.00% |

Table 3 shows the calculated FWHM value of the rocking curve by Meister and Weckermann's method and the new method in this article. The FWHMs of the neutron source and collimators in the table are commonly used in neutron scattering experiments. The relative deviations from the two methods are less than 3% in most cases. If the neutron source is fixed, then the smaller the FWHM of the collimators, the smaller will be the relative deviation. If the FWHM of the collimators is fixed, then the larger the FWHM of the neutron source, the smaller will be the relative deviation.

## 5. Experiment

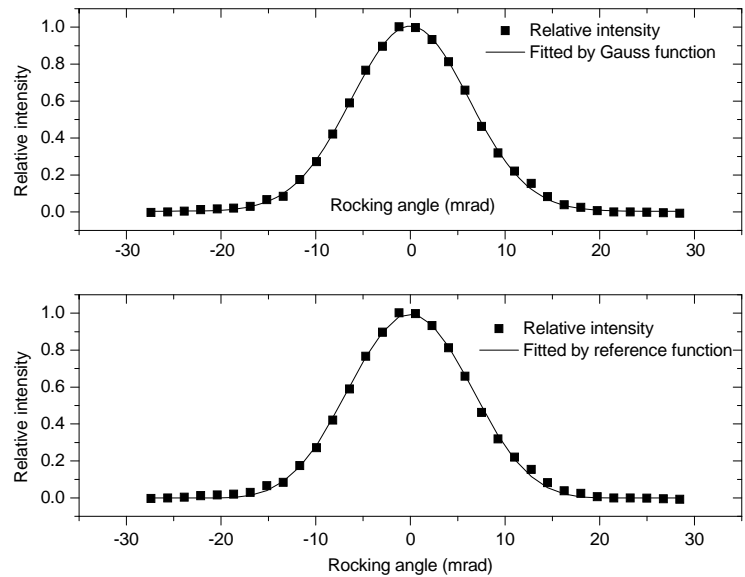

Fig.2. Rocking curve of the collimator ($\Gamma_1=\Gamma_2=36'$)

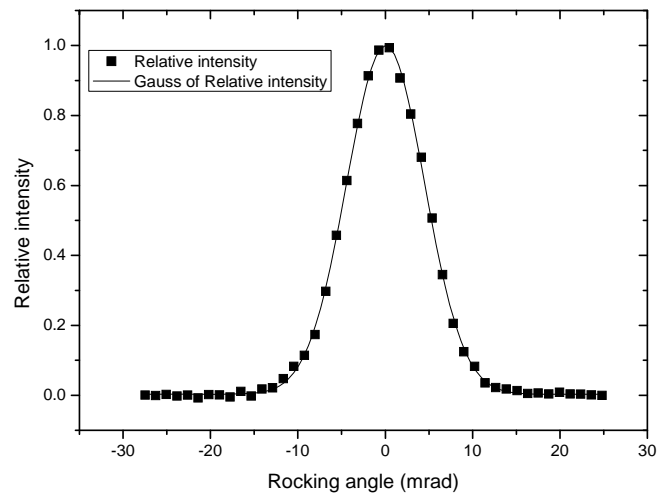

Fig.3. Rocking curve of the collimator ($\Gamma_1$=30′, $\Gamma_2$=20′)

In order to test the new method, two identical collimators (collimation angle $\Gamma_1$=$\Gamma_2$=36′) were measured using the SV30 Triple Axis Spectrometer at the China Institute of Atomic Energy, Beijing. The FWHM of the neutron source ($\Gamma_S$) is 43 mrad according to the instrument configuration. The experimental result for the rocking curve is shown in Figure 2. The top part of Figure 2 shows the curve fitted by a Gaussian function and the lower part shows the curve fitted by the reference function. Using the Gaussian function the FWHM ($\Gamma_{S+1+2}$) of the rocking curve is fitted to be 14.64±0.09mrad. The collimation angle is then calculated to be 36.12±0.25′using the Eq. (6). Its relative deviation is 0.3% compared with 36′. Using Meister and Weckermann's, the collimation angle is fitted to be 35.97±0.32′, and its relative deviation is 0.1% compared with 36′. Both uncertainties of the fitted values are less than 1% compared with 36′. The two results are consistent with the theoretical value within experimental uncertainties.

Another experiment was performed for two collimators with different collimation angles ($\Gamma_1$=30′, $\Gamma_2$=20′). The result is shown in Figure 3. By using the Gaussian function the FWHM ($\Gamma_{S+1+2}$) is fitted to be 10.49±0.06 mrad. The collimation angle $\Gamma_1$ is then calculated to be 30.14±0.26′using the Eq. (6), and its relative deviation is 0.5% compared with 30′. The uncertainty of the fitted value is less than 1% compared with 30′. One suggestion for this kind of configuration is that it is better to make the reference collimation angle less than the measuring collimation angle. In this way the fitted result could be more accurate.

## 6. Summary

A new method for analyzing the collimation angle of a neutron Soller collimator is described. A Gaussian distribution formula is used to describe the angular distribution function of the neutron source and Soller collimator. A relationship between the FWHM of the rocking curve and the collimation angle of the collimator is described. The experimental results show that the new function could be a good choice for fitting the experimental data.